\documentclass[english,aps,prl,amsmath,amssymb,twocolumn,letterpaper,citeautoscript,superscriptaddress,longbibliography]{revtex4-1}

\usepackage{mathtools} 

\usepackage[normalem]{ulem} 

\usepackage{comment}
\usepackage{amsmath}
\usepackage{physics}
\usepackage{graphicx}
\usepackage{babel}
\usepackage{amstext}
\usepackage{esint}
\usepackage[bookmarks=false,linkcolor=blue,urlcolor=blue,colorlinks,citecolor=blue]{hyperref}
\usepackage{pdfpages}	
\usepackage{pgffor}		

\makeatletter
\AtBeginDocument{\let\LS@rot\@undefined} 
\makeatother							 

\begin{document}

\title{Superconducting diode effect in quasi-one-dimensional systems} 

 \author{Tatiana de Picoli}

 \affiliation{Department of Physics and Astronomy, Purdue University, West Lafayette, Indiana 47907 USA}

 \author{Zane Blood}

 \affiliation{Department of Physics, Cornell University, Ithaca, New York 14850 USA}

\author{Yuli Lyanda-Geller}

\affiliation{Department of Physics and Astronomy, Purdue University, West Lafayette, Indiana 47907 USA}

 \author{Jukka I. V\"ayrynen}

 \affiliation{Department of Physics and Astronomy, Purdue University, West Lafayette, Indiana 47907 USA}

\date{\today}
\begin{abstract}
The recent observations of the superconducting diode effect pose the challenge to fully understand the necessary ingredients for non-reciprocal phenomena in superconductors. In this theoretical work, we focus on the non-reciprocity of  the critical current in a quasi-one-dimensional superconductor. 
We define the critical current as the value of the supercurrent at which the quasiparticle excitation gap closes (depairing). Once the critical current is exceeded,  the quasiparticles can exchange energy with the superconducting condensate, giving rise to dissipation. 
Our minimal model can be microscopically derived as a  low-energy limit of a Rashba spin-orbit coupled superconductor in a Zeeman  field. 
Within the proposed model, we explore the nature of the non-reciprocal effects of the critical current both analytically and numerically. 
Our results quantify how system parameters  such as spin-orbit coupling and quantum confinement 
affect the strength of the superconducting diode effect.  
Our theory provides a complementary description to   Ginzburg-Landau theories of the  effect. 
\end{abstract}
\maketitle

\textbf{\emph{Introduction.~}}
Since their discovery, diodes have played an important role in the development of new technologies.
Recently, the observation of non-reciprocity in the critical current of superconductors, known as the superconducting diode effect (SDE) \cite{harrington2009practical,vodolazov2018peculiar,ando2020observation}, has brought attention to this phenomenon for its  potential to achieve dissipationless electronics.
Following the initial observations, extensive work has been done to show the signature of SDE in different bulk materials \cite{wakatsuki2017nonreciprocal,2022arXiv220509276H,sundaresh21,bauriedl2021supercurrent, PhysRevLett.121.026601}. 
This diode effect  
was also observed and thoroughly
 studied in   Josephson junctions \cite{baumgartner2022supercurrent,diez2021magnetic,Mazur2022,baumgartner2022effect,PhysRevB.98.075430,https://doi.org/10.48550/arxiv.2211.10572,davydova2022universal,zhang2021general} (first in the context of the anomalous Josephson effect~\cite{PhysRevB.87.100506,PhysRevB.92.035428,PhysRevB.93.174502,2022arXiv221001037H,2023arXiv230101881B}) 
 and even in the absence of an applied magnetic field \cite{lin2021zero,wu2021realization,kokkeler2022field,scammell2021theory,Chiles2022}.

In general, non-reciprocity of the critical current occurs due to a broken inversion symmetry, which can be accomplished by an extrinsic or intrinsic mechanism. The first refers to the geometry of the system, 
the canonical example being 
an asymmetric superconducting ring
threaded by a magnetic flux~\cite{2014JETPL..99..169B,sundaresh21}. 
The SDE can also occur due to an intrinsic mechanism, for example, when one breaks the inversion of symmetry with spin-orbit coupling (SOC) \cite{yuan2021supercurrent,ilic2022theory,Daido2022,2022arXiv220315608K}. 
However, experimentally it can be a challenge to determine whether the non-reciprocity comes strictly from the intrinsic features of the system~\cite{sundaresh21}. 
Even theoretically, the exact minimal requirements for an intrinsic SDE remain unclear~\cite{agterberg2012magnetoelectric,LGinprep}. 
Most previous theoretical studies of intrinsic SDE have focused on using the Ginzburg-Landau theory (GL) \cite{yuan2021supercurrent,Daido2022,he2022phenomenological,daido2022intrinsic}, which is valid near the critical temperature $T\approx T_c$. While the phenomenological study of the SDE is able to help understand the effect near the critical temperature, a microscopic description of the phenomenon is still missing. 

\begin{figure}
    \centering
    \includegraphics{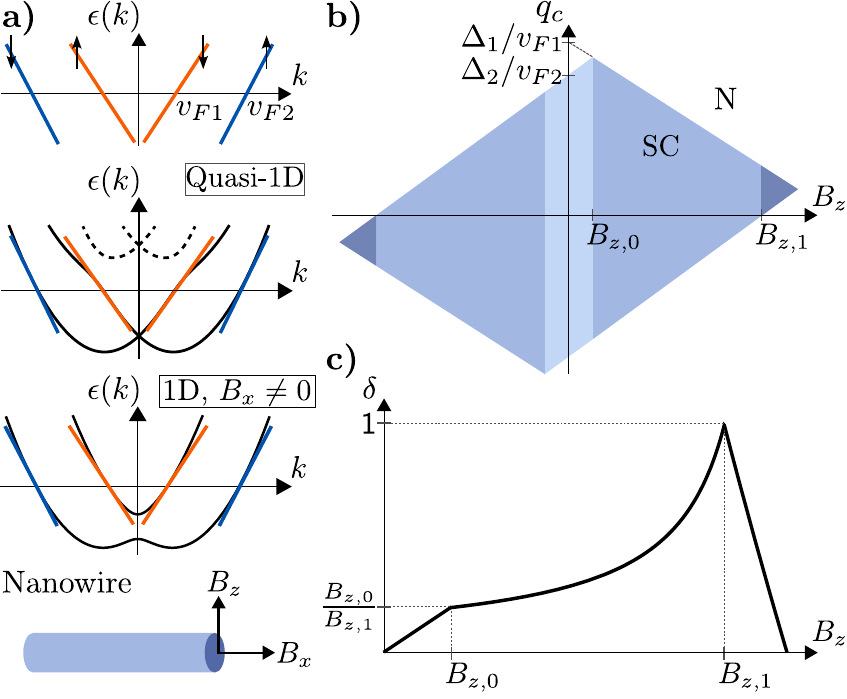}
    \caption{(a) \textit{top}: Linearized energy spectrum in the normal state, showing two helical bands with unequal Fermi velocities and opposite helicities. 
    The linearized model captures the Fermi level physics of   quasi-1D and 1D Rashba models (bottom).  
    In the quasi-1D case, hybridization 
    between the lower (solid) and upper (dashed) Rashba bands leads to unequal Fermi velocities. 
    In the purely 1D case the same can be achieved by applying a magnetic field $B_x$ along the wire. 
    (b) Phase diagram of superconducting (SC) and normal state (N) determined by the critical momentum $q_c$ as a function of the magnetic field $B_z$. The values of the magnetic field $B_{z,0}$ and $B_{z,1}$ delimit three different regions of the phase diagram. 
    (c) Quality factor $\delta$ versus magnetic field $B_z$ for the phase diagram shown in (b). The three regions of the phase diagram results in three different behaviors for the quality factor function.   }
    \label{fig:figure-1}
\end{figure}

In this work, we present a Bogoliubov-de Gennes model that describes the main mechanisms to achieve the intrinsic diode effect in uniform 1D superconductors. 
In 1D single-band regime, time-reversal invariant electronic systems can be generally described by a  Hamiltonian of two helical bands with opposite helicities.  
We show that unequal Fermi velocities of the two helical bands generically leads to the SDE, see Fig.~\ref{fig:figure-1}a. 
Microscopically, we show that this happens in 
Rashba systems
under quantum confinement or applied perpendicular magnetic field.
Generally, an applied supercurrent can be written as $I_S = \rho_s \hbar q$ where $\rho_s = en/2m$ is the superfluid stiffness (in terms of the 1D superfluid density $n$ and mass $m$) and $\hbar q$ is the Cooper pair momentum~\cite{tinkham2004introduction,PhysRevB.49.6841}. 
At low temperatures, when the superfluid density does not get appreciably modified by supercurrent, 
the study of non-reciprocity of the critical current $I_c = \rho_s q_c$ can be accomplished by calculating the critical momentum $q_c$ by using  the Cooper pair depairing condition~\cite{tinkham2004introduction,PhysRevB.49.6841,RevModPhys.34.667} (we set $\hbar = k_B = 1$ hereon). 
Focusing here on $s$-wave pairing, each helical band (labeled by $i$) can be treated independently and has a superconducting gap $\Delta_i$ at the Fermi level. 
 Qualitatively, Landau's criteria~\cite{landau1941theory} in the absence of magnetic field gives $q_{ci}=\Delta_i/v_{Fi}$ for the $i$-th helical band;  the critical momentum of the system  is then $q_c=\min_{i}\{q_{ci}\}$. 
 For two bands with opposite helicities, applying a magnetic field $B_z$ along the spin quantization axis will lower one $q_{ci}$ while increasing the other [see  Eq.~\eqref{critical_momentum}]. 
 Even with equal gaps, $\Delta_1=\Delta_2$, if the two bands have unequal Fermi velocities $v_{F1} \neq v_{F2}$, their critical momenta will become equal at a non-zero magnetic field  $B_{z} = B_{z,0}$, leading to  
 non-reciprocity. 
 This behavior is shown in Fig.~\ref{fig:figure-1}b,c.  
 For pairing $\Delta_i=\Delta$, the non-reciprocal behavior of the critical current, determined by the critical momentum, is fully explained by the difference of Fermi velocities, which carries information about inversion symmetry breaking.

\textbf{\emph{Low-energy model.~}}
To investigate the mechanisms responsible for the appearance of an intrinsic non-reciprocal behavior of the critical current, we investigate a low-energy  minimal model. 
We propose a Hamiltonian of two helical bands,  
\begin{align}
    H= \frac{1}{2}  \sum_k C^\dagger_k\mathcal{H}_{BdG}C_k,
\end{align}
where $C_k$ is an eight component Nambu spinor defined by   $C_k=(\begin{array}{cc}
C_{k1} & C_{k2}\end{array})^T$ with $C_{ ki}=(\begin{array}{cccc}
c_{ k+qi\uparrow} & c_{ k+qi\downarrow} & c_{  -k+qi\downarrow}^{\dagger} & -c_{  -k+qi\uparrow}^{\dagger }\end{array})^T$. In this representation, $\uparrow,\downarrow$ are spins (or pseudo-spins) of our system and the subscript $i$ is the label for each helical band. 
The Bogoliubov-de Gennes (BdG) Hamiltonian is given by 
$\mathcal{H}_{BdG} = \text{diag}(\mathcal{H}_{k1}, \mathcal{H}_{k2})$ with  ($\chi_i = -(-1)^i$) and
\begin{align}
&\mathcal{H}_{ki}\!=\! v_{Fi}\left(\chi_i  k\sigma_{z} \! - \! k_{Fi}\right)\tau_{z}+\chi_i v_{Fi}q\sigma_{z}\!+\!\Delta_{i}\tau_{x}\!+\!\frac{g_{i}\mu_B}{2}\vec{B}\cdot\vec{\sigma}.
\label{hamilt-toy-model}
\end{align} 
In this effective model, each helical band 
is allowed to 
have in general an independent Fermi velocity $v_{Fi}$, Fermi momentum $k_{Fi}$, s-wave (intra-band) pairing gap $\Delta_{i}$, and g-factor $g_{i}$, 
while experiencing the same applied magnetic field $\vec{B} $ ($\mu_B$  denotes the Bohr magneton). We have linearized the dispersion, focusing on low energies near the Fermi surface (see Fig.~\ref{fig:figure-1}a). 
The Pauli matrices $\sigma_{x,y,z}$ and $\tau_{x,y,z}$ act on the spin and particle-hole spaces respectively.
The parameter $q$ is the Cooper pair momentum of the superconductor and determined by the externally applied supercurrent.  Considering proximity-induced superconductivity at low temperatures, we are able to relate the Cooper pair momentum $q$ to the applied supercurrent as 
$q\propto I_S$ in the first approximation.  
The analysis of non-reciprocity of the critical current can be performed by studying the behavior of the critical Cooper pair momentum $q_c$ as a function of applied magnetic field. 

From the above Hamiltonian~\eqref{hamilt-toy-model}, we find the energy cost $E_{\sigma i}(k)$ to create an excited above-gap quasiparticle of spin  $\sigma=\uparrow,\downarrow=+,-$ and momentum $k$  in the band $i$. For an applied magnetic field  $\vec{B}=B_{z}\hat{z}$, this energy becomes 
\begin{align}
E_{\sigma i}(k) & \!=\!\sigma\!\left(\!\frac{g_{i}\mu_B}{2}B_{z}+\chi_i  qv_{Fi}\!\right)\!+\!\!\sqrt{\Delta_{i}^{2}+ \! \left(k\! - \! \chi_i \sigma k_{Fi}\right)^{2} \! v_{Fi}^{2}}.
\label{spectrum-toy-model}
\end{align}

Assuming for the moment $B_z$ and $q$ such that this energy cost is positive, the excitation energy of a quasiparticle will be the smallest at the Fermi momentum $k_{Fi}$. 
This energy cost can increase or decrease by tuning the applied magnetic field $B_z$ and momentum $q$. 

The critical momentum $q_c$ of the system is the specific momentum for which $E_{\sigma i}(k=\chi_i \sigma k_{Fi})=0$, i.e., there is no energy cost to create a quasiparticle excitation \cite{ginzburg2009theory,RevModPhys.34.667,landau1941theory}. For an applied current larger than the critical one, we expect the system to be in the normal phase (N) instead of the superconducting one (SC). Therefore, we focus our description for $q\leq q_c$.
From the dispersion \eqref{spectrum-toy-model} we find that the critical momentum for each helical band is a linear function of the magnetic field
\begin{equation}
q_{ci}^{\pm}=\frac{ - \chi_i \frac{1}{2}g_{i}\mu_BB_{z}\pm\Delta_{i}}{v_{Fi}} \,,
\label{critical_momentum}
\end{equation} 
where, the superscript $\pm$ labels the direction of the applied supercurrent. 
The critical momentum of the two-band system is then, $q_c^{\pm} = \pm \min_{i=1,2} |q_{ci}^{\pm }|$.

The non-reciprocal behavior occurs when, for a fixed magnetic field, the absolute value of the critical current is different in the positive and negative directions. In terms of the critical momentum the non-reciprocity condition translates to $|q_{c}^{+}|\neq |q_{c}^{-}|$. 
To better understand how the physical parameters contribute to the superconducting diode effect, we can define a quality factor of the critical current as
\begin{equation}
 \delta  =\frac{|q_{c}^{+}|-|q_{c}^{-}|}{|q_{c}^{+}|+|q_{c}^{-}|}.
 \label{quality-factor}
\end{equation}
The phase diagram in Fig.~\ref{fig:figure-1}b shows the phase separation between normal and superconducting phase determined by the four components of the critical momentum \eqref{critical_momentum}. We define by $B_{z,0}$ ($-B_{z,0}$) the magnetic field in which the critical momentum of different helical bands first cross, i.e, $q_{c1}^{+}=q_{c2}^{+}$ ($q_{c1}^{-}=q_{c2}^{-}$). Another characteristic value 
is  the magnetic field $B_{z,1}$ in which the critical momentum $q_{ci}^{-}$ changes sign. 
Without loss of generality, we assume $\Delta_2/v_{F2}<\Delta_1/v_{F1} $.  
For this choice of parameters, the explicit forms of $B_{z,0}$ and $B_{z,1}$ are found to be 
\begin{equation}
    B_{z,0}=\frac{\bar{\Delta}_1\bar{v}_{F2}-\bar{\Delta}_2\bar{v}_{F1}}{\bar{v}_{F1}+\bar{v}_{F2}}, 
    \quad B_{z,1}= \min \{ \bar{\Delta}_1, \bar{\Delta}_2 \} , 
    \label{eq:Bzbar}
\end{equation}
where  we define $\bar{v}_{Fi}=v_{Fi}/(\frac{1}{2}g_{i}\mu_B)$, $\bar{\Delta}_i=\Delta_i/(\frac{1}{2}g_{i}\mu_B)$. 
In  Fig.~\ref{fig:figure-1}c we show the behavior of the quality factor as a function of the magnetic field when $0 < B_{z,0} < B_{z,1}$. 
In general, we have a linear increase in the quality factor for small magnetic field, i.e., $\delta=B_z/B_{z,1}$ for $|B_z|<|B_{z,0}|$. For a larger magnetic field, the behavior of the quality factor is dependent on the particular choice of parameters. 
For the particular case shown in Fig.~\ref{fig:figure-1}c, where $B_{z,0}/B_{z,1}\ll 1$ the quality factor can be approximated by 
\begin{equation}
\delta\approx \frac{B_z}{|B_z|}\frac{B_{z,0}}{-|B_z|+B_{z,0}+B_{z,1}},\ \ (B_{z,0}<|B_z|<B_{z,1}) ,
    \label{quality-factor-multichannel}
\end{equation}
reaching its maximum value 1 at $B_{z,1}$  when $|q_{c}^{-}|=0$. 
The critical current becomes reciprocal and diode effect disappears in the limit ${B_{z,0}\to 0}$. In this limit, the quality factor~(\ref{quality-factor}) becomes ill-defined at the critical field $B_z = B_{z,1}$, but for any $B_z < B_{z,1}$ such that $|B_{z,1} - B_z | \gg B_{z,0}$,  Eq.~(\ref{quality-factor-multichannel}) yields  a quality factor that vanishes as $\delta \propto B_{z,0}$ in the reciprocal limit. 
The ratio $ B_{z,0} /  B_{z,1}$, when small, is a characteristic measure for weak diode effect.

\textbf{\emph{Self-consistent gap. }}  
So far we focused the analysis of the transition between superconducting to normal phase only on the critical Cooper pair momentum $q_c$. One could argue that $|q|>|q_c|$ is not a sufficient condition to ensure that the system is in the normal phase, i.e., that superconductivity could survive even in a gapless system. To study the practicability of such gapless superconductivity, we calculate the pairing potential self-consistently \cite{de2018superconductivity}.
This 
calculation also allows us to extend the low-energy model described 
to finite temperature.
For one helical band, i.e., choosing  subsystem $i$ of \eqref{hamilt-toy-model}, the self-consistency consists of solving for $\Delta_i \equiv\Delta_i(q, T)$ the equation 
\begin{equation}
    1\!=\!V_i\int_{-k_D}^{k_D} \! \frac{dk}{2\pi}\frac{ 1-f[E_{\uparrow i}(k)]-f[E_{\downarrow i}(-k)]}{2\sqrt{\Delta_i^{2}(q, T)+(k-\chi_ik_{Fi})^{2}v_{Fi}^{2}}}, 
    \label{self-consistency}
\end{equation}
where $f[E_{\sigma i}(k)]$ is the Fermi-Dirac distribution, $E_{\sigma i}(k)$ is the dispersion \eqref{spectrum-toy-model} calculated at $B_z=0$, $V_i/v_{Fi}$ is the dimensioneless pairing interaction strength and $k_D$ is the Debye wave vector, providing a UV cutoff. 
 
\begin{figure}[tb]
\centering

\includegraphics[width=0.9\columnwidth]{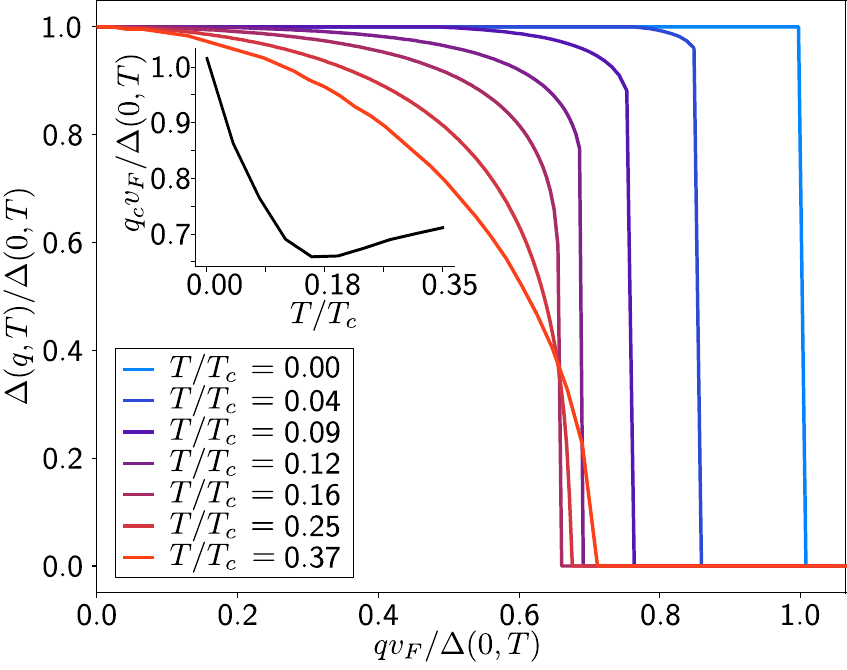}
\caption{Self-consistently calculated gap $\Delta(q,T)$ [in units of $\Delta(0,T)$] versus Cooper pair momentum $q$ [in units of $\Delta(0,T)/v_F$].
\textit{Inset:}~The critical Cooper pair momentum $q_c(T)$ versus the temperature $T$ normalized by $T_c=\Delta(0,0)/(1.76k_B)$. Here, $q_c(T)$ is defined as the smallest $q$ such that $\Delta(q,T) = 0$. 
This shows that at non-zero temperature, 
Eq.~\eqref{critical_momentum} can be approximately used with a temperature-dependent gap $\Delta(T)$ multiplied by a weakly temperature-dependent coefficient $q_c(T) / \Delta(0,T) $. 
The shown results are with $B_z = 0$; a non-zero $B_z$ adds linearly to $q$, see Eq.~\eqref{critical_momentum}.
}
\label{fig:self-consistent}
\end{figure}

In Fig.~\ref{fig:self-consistent} we show $\Delta(q, T)$ versus Cooper pair momentum $q$ plot for different values of temperature, obtained by solving Eq.~\eqref{self-consistency} numerically.
For $T=0$ we find that $\Delta(q, 0)$ is constant for Cooper pair momentum $q$ below the critical one. For $q>q_c$, Eq.~\eqref{self-consistency} has no solution, showing that the critical momentum found, Eq.~\eqref{critical_momentum}, is the correct threshold to determine SC to N transition in our 1D helical model. 
A non-zero $B_z$ will only shift $q_{ci}$ linearly, as described by  Eqs.~\eqref{spectrum-toy-model}-\eqref{critical_momentum}.

\textbf{\emph{Microscopic models.~}} 
Up to now, we have  described an effective low-energy model that shows non-reciprocal phenomena and the mechanisms that allow the existence of the SDE. To complete our discussion, it is important to understand microscopically how to achieve unequal Fermi velocities between two helical bands. 
Here we describe two Rashba systems that, in the low-energy limit, can be well described by our minimal model~\footnote{We expect similar physics in gate-defined wires in HgTe quantum wells~\cite{PhysRevX.3.031011}}. 

\textbf{\emph{Quasi-1D Rashba wire.~}} 
We start by  considering a quasi-1D Rashba nanowire in the presence of a Zeeman field, described by the normal-state Hamiltonian~\cite{Alicea2012}, 
\begin{equation}
H=\!\!\int\!\! dx\hat\Psi^{\dagger}(x) \left( \mathcal{H}_{0}+\mathcal{H}_{R}+\mathcal{H}_{Z} \right)\!\hat\Psi(x), 
\label{hamiltonian-multichannel-rashba}
\end{equation}
with $\mathcal{H}_{0}  = -\frac{1}{2m} \partial_{x}^{2}-\mu+E_{0}\Sigma_{z}$ and 
\begin{equation}
\!\mathcal{H}_{R} \! = -i\alpha\partial_{x}\sigma_{z}+\eta\sigma_{x}\Sigma_{y}\,, \quad
\!\mathcal{H}_{Z}\!= \frac{1}{2}g\mu_B\vec{B}\cdot\vec{\sigma} ,\label{magnetic_term}
\end{equation}
where $\hat\Psi(x)=\left(\begin{array}{cccc}
\hat\psi_{1\uparrow}(x) & \hat\psi_{1\downarrow}(x) & \hat\psi_{2\uparrow}(x) & \hat\psi_{2\downarrow}(x)\end{array}\right)^{T}$ and $\Sigma_{x,y,z}$ are  Pauli matrices  that act on the transverse degree of freedom. Here, we consider the two lowest-energy transverse modes labeled by $i=1,2$. 
The Hamiltonian $\mathcal{H}_0$ describes the kinetic energy and confinement gap $2 E_0 \propto 1/W^2$ between transverse bands. The Rashba Hamiltonian $\mathcal{H}_R$ is written in terms of $\alpha$ and $\eta \propto 1/W$ denoting the spin-orbit couplings respectively along and perpendicular to the wire. 
The parameters $\eta$ and $E_0$ depend on the width $W$ of the wire and the specific confining potential, see Ref.~\cite{2017PhRvB..96l5416P}. Our analysis, however, is completely independent regarding the specific forms of these parameters. 

We first analyse $\mathcal{H}_0+\mathcal{H}_R$ for the range of energy where only the lowest energy  transverse channel is occupied. This Hamiltonian commutes with the pseudo-spin operator $\sigma_z\Sigma_z$, therefore it is convenient to label the energies with  $\sigma_z\Sigma_z$ eigenvalues $\pm 1$. The dispersion of the lowest transverse mode is given by
\begin{equation}
\epsilon_{\pm}(k)=\frac{k^{2}}{2m}-\mu-\sqrt{(E_{0}\pm\alpha k)^{2}+\eta^{2}}.
\end{equation}
From the dispersion, we find two positive Fermi momenta, $k_{F1,2}$, where $k_{Fi}$ obeys $\epsilon_{\pm}(k_{Fi})=0$. 
We also find the Fermi velocities
\begin{equation}
v_{Fi}  =\frac{k_{Fi}}{m}-\frac{\alpha(\alpha k_{Fi}\pm E_{0})}{\sqrt{(\alpha k_{Fi}\pm E_{0})^2+\eta^{2}}},\label{eq:velocity}
\end{equation} 
around $k_{F1}$ (-) and $k_{F2}$ (+).

In order to study the effects of weak magnetic field and proximity-induced superconductivity near the Fermi momenta, we  linearize the dispersion by writing the field operator $\hat\Psi(x)$ as a superposition of left and right-movers for each  pseudo-spin subbands, 
\begin{align}
    \!\!\hat \Psi(x)\!=&\!\left[\hat\psi_{R\uparrow}(x) \sigma_x e^{ik_{F1}x}\!+\!\hat\psi_{L\downarrow}(x)e^{-ik_{F1}x}\right]\!\!\phi_1\nonumber\\
    \!+&\!\left[\hat\psi_{R\downarrow}(x)e^{ik_{F2}x}\!+\!\hat\psi_{L\uparrow}(x)\sigma_xe^{-ik_{F2}x}\right]\!\!\phi_2,
\label{decomposition-multichannel}
\end{align}
where $\phi_i=\left( 
\begin{array}{cccc}
 i\sin\frac{\theta_{i}}{2} & 0 & 0 & \cos\frac{\theta_{i}}{2}\end{array} \right)^T$ and $\theta_i=\arccos[\pm{\alpha^{-1}}\left(v_{Fi}-{k_{Fi}}/{m}\right)]$ with $+,-$ for $i=1,2$, respectively.
We apply \eqref{decomposition-multichannel} to  the 
Hamiltonian \eqref{hamiltonian-multichannel-rashba} for $\vec{B}=B_{z}\hat{z}$ to obtain an effective model for the quasi-1D nanowire in a perpendicular magnetic field. 
To obtain the low-energy description near the Fermi points, we assume that the components $\psi_{R(L)\sigma}(x)$ vary slowly in space allowing us to neglect terms $\partial^2_x\psi_{R(L)\sigma}(x)$.  
Likewise, fast oscillating terms $\propto e^{\pm i(k_{Fi}+k_{Fj})x}$ are also neglected~\cite{2017PhRvB..96g5404V}. 
We find  the linearized dispersion of the nanowire in the normal phase, 
\begin{align}
    \epsilon_{\sigma i}(k) & \!=\!\sigma v_{Fi}\left(k-\sigma\chi_i  k_{Fi}\right)+\sigma\frac{g_{i}\mu_B}{2}B_z,
\end{align}
where $g_{i}=g\cos\theta_i$. 
Finally, we include 
proximity-induced superconductivity with intrachannel pairing $\Delta e^{-2iqx}\sum_{i=1}^{2}\hat\psi_{i\uparrow}\hat\psi_{i\downarrow}$ with Cooper pair momentum $q$.
Linearizing the pairing term by 
substituting \eqref{decomposition-multichannel} into it, 
we are able to write the quasi-1D Rashba system using our minimal model Hamiltonian  \eqref{hamilt-toy-model}. 
Here we find induced gaps $\Delta_1=\Delta_2=\Delta$ at the two Fermi momenta $k_{F1,2}$, respectively.   

To understand the behavior of the quality factor $\delta$ in the quasi-1D case, we consider the limit $E_0\gg\alpha^2m,\eta,\mu$. 
In this regime, the energy difference between transverse bands is large, so the upper bands are unoccupied, but the hybridization $\eta$ of the bands  
will change the Fermi velocities $v_{F1,2}$ 
by a small factor. 
To show the effects of small transverse coupling we expand the velocities $v_{Fi}$ and g-factor $g_i$ in powers of $\eta$. Plugging this expansion into the expression for $B_{z,0}$ given by Eq.~\eqref{eq:Bzbar}, we find   $B_{z,0}\approx 2(m\alpha^2 \eta^2/E_0^3)(\alpha /v_{F}) B_{z,1}$ and $B_{z,1} \approx \Delta / (\frac{1}{2} g \mu_B)$, where $v_F=\sqrt{2E_{0}/m}$. 
Thus, non-reciprocity arises in high order in spin-orbit coupling, stemming from weak hybridization of the transverse modes~\cite{Note1}. 

\textbf{\emph{Purely 1D Rashba wire with $B_x$.~}} 
As seen above, in the purely 1D model  ($W\to0,E_0 \to \infty$), 
the critical current becomes reciprocal ($B_{z,0}\to 0$). 
However, even in this case we can  induce  non-reciprocity  by applying a transverse magnetic field which will lead to unequal velocities of the inner and outer Rashba modes, see Fig.~\ref{fig:figure-1}a. 

In the 1D limit the energy splitting  $E_0$  between transverse bands is  large and  we can  
project the Hamiltonian Eq.~\eqref{hamiltonian-multichannel-rashba} to the states with $\Sigma_z = -1$.  
Now the Hamiltonian commutes with $\sigma_z$ (eigenvalues $\sigma = \pm 1$) and 
the energy dispersion gives
equal Fermi velocities $v_F = \sqrt{2\mu/m+\alpha^{2}} $ for the inner and outer Rashba modes~\cite{PhysRevB.92.014509}. 
In this case, there is no non-reciprocity~\footnote{The spin-orbit coupling can be gauged out in this case~\cite{LyandaGeller1994}.}. 
Next, we consider an additional component of the magnetic field in Eq.~\eqref{magnetic_term} as $B_x\sigma_x$,
that acts in a similar way to the coupling $\eta$ by breaking the conservation of spin~\cite{PhysRevLett.109.226804}. To understand how the transverse magnetic field changes the velocities of the helical bands, we will treat this term perturbatively in the superconducting phase. First, we note that in the normal phase, the transverse magnetic field opens a gap at $k=0$, affecting the inner helical band with smaller Fermi momentum [$k_{F1}=(v_F-\alpha )m$] 
while presenting a negligible effect on the outer helical band with $k_{F2}$, as long as $\frac{1}{2} g \mu_B B_x \ll m \alpha v_F $.
In the presence of proximity-induced superconductivity, the helical band around $k= k_{F1}$ (and similarly for $k=- k_{F1}$)
can be described as 
$\mathcal{H}_{k1}=\text{diag}( h_{k\uparrow}, h_{k\downarrow})$, where $h_{k\uparrow} =B_{z}+qv_{F1}+\Delta\tau_x$  and $h_{k\downarrow}\approx-2v_{F1}k_{F1} \tau_z $. 
By finding the eigenstates in the proximity of the Fermi level $\pm k_{F1}$, we can calculate the energy correction due to the applied perturbation $B_x$.
 We find $g_1=g+\delta g$ and $v_{F1}=v_{F}+\delta v_{F}$, where $\frac{\delta v_{F}}{v_{F}}=\frac{\delta g}{g}=-\left(\frac{1}{4}\frac{g\mu_{B}B_{x}}{{k_{F1}v_{F}}}\right)^{2}$, resulting in $v_{F2}>v_{F1}$, $g_2>g_1$ and $\Delta_1=\Delta_2$, thus leading to non-reciprocal critical current. 
In this case, we find   $B_{z,0} \approx \frac{1}{2}\left(\frac{1}{4}\frac{g\mu_{B}B_{x}}{{k_{F1}v_{F}}}\right)^{2}B_{z,1}$ while $B_{z,1}\approx \Delta / (\frac{1}{2} g \mu_B) $, 
which implies the behavior of the quality factor as seen in    Fig.~\ref{fig:figure-1}c. 

\textbf{\emph{Discussion.~}} 
We showed that helical bands with a Fermi velocity difference $\delta v_F$ give rise to critical current non-reciprocity with the size of the effect quantified by $B_{z,0}/B_{z,1}\approx \delta v_F / v_F$ [see below Eq.~(\ref{quality-factor-multichannel})]. 
This form shows that the intrinsic superconducting diode effect is generally small in ordinary metals: the denominator is the Fermi velocity which increases with electron density whereas the numerator is the difference of Fermi velocities and typically (at most) of the order of the spin-orbit velocity, independent of the density. 
In a quasi-1D system of width $W$, we found that $\delta v_F \sim \alpha m\alpha^{2} \eta^{2}/E_0^3$ arises   from a transverse spin-orbit coupling $\eta \sim \alpha /W $. 
Since $E_0 \sim 1/(m_{\perp} W^2) $, we see that $\delta v_F \propto (W / l_{\alpha})^4$ increases with the width of the system. Here, we introduced the Rashba length $l_{\alpha}=1/(m\alpha)$ and assumed
isotropic  
effective mass $m_{\perp} \approx m$. 
This quasi-1D result is valid in the limit of small width,
$W \ll l_{\alpha}$. 
The opposite limit of large $W / l_{\alpha}$, is for general chemical potential a complex problem due to multiple bands and Fermi points. Nevertheless, in the low density case $\mu \ll m \alpha^2$ there are only two Fermi points, and we  obtain a simple result $\delta v_F \sim \alpha E_{0} / (m\alpha^{2}) \propto (l_{\alpha} / W)^2 $ by treating $E_0$ perturbatively. 
Thus, in low-density, clean, Rashba wires the non-reciprocity is a non-monotonic function of the wire width, with a maximum 
at $W$ of order the Rashba spin-orbit length, estimated to be of order $100\mathrm{nm}$~\cite{PhysRevB.71.205328,PhysRevLett.98.266801,PhysRevB.81.155449,PhysRevB.97.041303}. 
We emphasize however that this simple consideration is only valid in the low-density single-band regime and ignores disorder, Dresselhaus spin-orbit coupling, and mass  anisotropies. 
Further studies are needed to make quantitative  predictions for optimizing the strength of intrinsic non-reciprocity in Rashba systems.

\begin{acknowledgments} 
\textbf{\emph{Acknowledgments.~}} 
We thank  Leonid Rokhinson and  Ananthesh Sundaresh for a related collaboration and discussions, and Leonid Glazman for valuable remarks. 
TdP acknowledges the Graduate School for support under the Ross Fellowship Program. 
 ZB was supported by NSF REU grant PHY-1852501 and the ALPHA collaboration. 
This material is based upon work supported by 
the U.S. Department of Energy, Office of Basic Energy Sciences, Division of Materials Sciences and Engineering under Award DE-SC0010544 (Y.L-G) 
and by 
the  Office of the Under Secretary of Defense for Research and Engineering under award number FA9550-22-1-0354 (JIV). 

This work was performed in part at Aspen Center for Physics, which is supported by National Science Foundation grant PHY-1607611. 
\end{acknowledgments}

\bibliography{refs}

\end{document}